\documentclass[%
reprint,
superscriptaddress,
showpacs,%
nofootinbib,
nobibnotes,
amsmath,amssymb,
aps,
pre,
floatfix,
]{revtex4-1}

\usepackage[utf8]{inputenc}
\usepackage[T1]{fontenc}
\usepackage[english]{babel}
\usepackage{graphicx}
\usepackage{color}
\usepackage{hyperref}

\hypersetup{colorlinks = true, allcolors = blue}

\usepackage{soul} 

\newcommand{\widfig}[2]{\includegraphics[width=#1\columnwidth]{#2}}
\newcommand{\fulfig}[1]{\includegraphics[width=\textwidth]{#1}}

\newcommand{\mat}[1]{\mbox{\boldmath{$\mathrm{#1}$}}}
\newcommand{\identmat}{\mat I}
\newcommand{\diagmat}{\mathop{\mathrm{diag}}}
\newcommand{\clutim}{\mathcal{T}}

\newcommand{\ddip}{d^{\,\text{d}}}
\newcommand{\dpik}{d^{\,\text{p}}}
\newcommand{\dful}{d^{\,\text{f}}}
\newcommand{\dcom}{d^{\,\text{c}}}
\newcommand{\subsub}{\text{sub-sub}}
\newcommand{\hubsub}{\text{hub-sub}}
\newcommand{\dash}{\text{-}}

\newcommand{\variance}{\mathop{\mathrm{Var}}}

\newcommand{\meye}[1]{\mat e_{#1,#1}}
\newcommand{\mzergen}[2]{\mat z_{#1,#2}}
\newcommand{\monegen}[2]{\mat d_{#1,#2}}

\newcommand{\mzer}[1]{\mzergen{#1}{#1}}

\newcommand{\mone}[1]{\monegen{#1}{#1}}
\newcommand{\vone}[1]{\monegen{#1}{1}}
\newcommand{\vtone}[1]{\monegen{1}{#1}}

\newcommand{\neweps}[1]{\textbullet{}~\mbox{#1:~}}
\newcommand{\subtwo}{\ensuremath{_{2}}}

\begin{document}

\title{Variety of regimes of star-like networks of H\'{e}non maps}

\author{Pavel V. Kuptsov}\email[Corresponding author. Electronic
address:]{p.kuptsov@rambler.ru}%
\affiliation{Institute of electronics and mechanical engineering, Yuri
  Gagarin State Technical University of Saratov, Politekhnicheskaya
  77, Saratov 410054, Russia}%

\author{Anna V. Kuptsova}%
\affiliation{Institute of electronics and mechanical engineering, Yuri
  Gagarin State Technical University of Saratov, Politekhnicheskaya
  77, Saratov 410054, Russia}%

\pacs{05.45.-a, 05.45.Xt, 05.45.Pq, 89.75.Hc}


\keywords{Star-like networks, Complex dynamical networks,
  Synchronization, Wild multistability, Remote synchronization}

\date{\today}

\begin{abstract}
  In this paper we categorize dynamical regimes demonstrated by
  star-like networks with chaotic nodes. This analysis is done in view
  of further studying of chaotic scale-free networks, since a
  star-like structure is the main motif of them. We analyze star-like
  networks of H\'{e}non maps. They are found to demonstrate a huge
  diversity of regimes. Varying the coupling strength we reveal chaos,
  quasiperiodicity, and periodicity. The nodes can be both fully- and
  phase-synchronized. The hub node can be either synchronized with the
  subordinate nodes or oscillate separately from fully synchronized
  subordinates. There is a range of wild multistability where the zoo
  of regimes is the most various. One can hardly predict here even a
  qualitative nature of the expected solution, since each perturbation
  of the coupling strength or initial conditions results in a new
  character of dynamics.
\end{abstract}

\maketitle

\section{Introduction}

Complex dynamical networks with scale-free coupling structure attract
a lot of interest as models for a large variety of natural
systems~\cite{Boccaletti2006175,CxNetwTopDynSyn2002}. The name
``scale-free'' for these networks appears because their node degree
distributions have power law shapes. As a result a small number of
nodes hold a major balk of links while the rest of nodes have few
connections~\cite{ScaleFreeNetw}.

One of the main questions arising in studying of these networks is the
type and conditions of synchronization~\cite{Boccaletti2006175,
  Arenas200893, osipov2007synchronization, golubitsky2015recent}. The
synchronization can be full or only the phases of node oscillators can
be synchronized; it can involve the whole bunch of nodes or the nodes
can form synchronized clusters~\cite{CxNetwTopDynSyn2002,
  SyncGraphTopol2005,
  arenas2006synchronization}. Papers~\cite{wang2002synchronization,
  SyncScaleFree2005, osipov2007synchronization, Arenas200893} are
devoted to an analysis of synchronization conditions,
works~\cite{JalanAmritkar2003, JalanAmritkar2005} investigate the
formation of synchronization clusters while in
Ref.~\cite{ImpLeadClSync2015} the impact of presence of a leader on
the cluster synchronization is recovered. Paper~\cite{RemSyn2}
investigates so called remote synchronization when nodes can get
synchronized even being connected indirectly though intermediate
ones. Also this regime is studied in Refs.~\cite{JalanAmritkar2003,
  JalanAmritkar2005} being called driven synchronization.

Authors of Ref.~\cite{Wang20101464} study scale-free networks with
fractional order oscillators. Paper~\cite{wang2002pinning}
investigates the control of a scale-free dynamical network by applying
local feedback injections to a fraction of network nodes. Covariant
Lyapunov vectors~\cite{CLV2012} and their nonwandering predictable
localization is studied in Ref.~\cite{NWL2014}.

The feature specific for scale-free networks as well as for other
complex dynamical networks is
multistability~\cite{Roxin2005,Angeli2004,Angeli2009398}.  It is well
known that the dynamics of multistable systems can be amazingly
rich~\cite{Feudel2008,Pisarchik2014167}. It occurs when the number of
attractors is very high, and their basins have fractal boundaries that
are highly interwoven. In this case the dynamics is very sensible to
the initial state: even tiny perturbation results in arriving at new
regime. Moreover, the ranges of existence of particular attractors can
be narrow so that the qualitative behavior of the system can change
dramatically when its parameters are slightly
varied~\cite{Feudel2008}. For scale-free networks this type of
behavior was reported in Ref.~\cite{NWL2014}. We suggest to refer to
this type of dynamics as \emph{wild multistability} to distinguish it
form the plain case, when one can easily locate basins of the required
attractors and put the system there to observe the expected behavior.

Wild multistability as well as other dynamical phenomena demonstrated
by dynamical networks still requires an exhaustive study. In
particular the interesting question is to reveal what features are the
same with other systems and what additionally emerge due to the strong
inhomogeneity of networks. For scale-free networks the this study can
be started from the consideration of their simplest and regular
representatives which are \emph{star-like structures}. These
structures consist of the hub node connected with all other nodes and
subordinate nodes that have only one connection with the
hub. Star-like structures are the main motifs of scale-free networks,
their building blocks.

Chaotic synchronization of oscillator networks with star-like
couplings is considered in Ref.~\cite{Pecora98}. Formation of
synchronized clusters in such networks is studied in
Ref.~\cite{ClustSyncStar2008} and a sufficient condition about the
existence and asymptotic stability of a cluster synchronization
invariant manifold is derived. Paper~\cite{RemSyn1} considers phase
synchronization of the subordinates when the hub is not synchronized
with them. This is called remote synchronization.

Thus, to pave the way to understanding the dynamics of scale-free
networks, one has to reveal first the details of dynamics of systems
with star-like coupling. This is the main motivation for the present
paper. We consider H\'{e}non maps. This map is a canonical model that
is sufficiently simple on the one hand and exhibit the same essential
properties as much more ``serious'' chaotic system on the other
hand~\cite{Henon1976}. In particular, this map is time-reversible, and
the coupling between network nodes is introduced in a way preserving
this property. Altogether, we build a catalog of regimes of
time-reversible star-like network of H\'{e}non maps observed at
different coupling strengths.  The most of the regimes are found to be
independent on the network size. However there is an area of wild
multistability where the dynamics does depend on the number of
nodes. Many different attractors exist here within a narrow ranges. As
a result, one can hardly predict even a qualitative nature of the
expected solution, since each perturbation of coupling strength or
initial conditions leads to a new character of dynamics.

The outline of the paper is the following. First we introduce the
model system. Then we discuss the numerical criteria for detecting
various synchronization regimes. After that the registered regimes are
represented and discussed. Finally, the ranges of stability of some of
the regimes are derived.

\section{Model system}

We consider a network of H\'{e}non maps introduced in
Ref.~\cite{NWL2014} as a generalization of the H\'{e}non chain from
Ref.~\cite{PolTor92a}:
\begin{equation}
  \label{eq:netw_henon}
  \begin{gathered}
    x_n(t+1)=\alpha-[x_n(t)+\epsilon h_n(t)]^2+y_n(t),\\
    y_n(t+1)=\beta x_n(t),\\
    h_n(t)=\sum_{j=1}^N \frac{a_{nj}}{k_n}x_j(t)-x_n(t),\;
    k_n=\sum_{j=1}^Na_{jn}.
  \end{gathered}
\end{equation}
Here $N$ is the number of network nodes, $t=0,1,2\ldots$ is discrete
time, $a_{nj}\in \{0,1\}$, $a_{nj}=a_{jn}$, $a_{nn}=0$ are the
elements of the $N\times N$ adjacency matrix $\mat A$, and $k_n$ is
degree of the $n$th node, i.e., the number of its
connections. $\alpha=1.4$ and $\beta=0.3$ are the parameters
controlling local dynamics, and $\epsilon\in [0,1]$ is the coupling
strength. Recall that the H\'{e}non map is time-reversible. The
coupling is introduced in a way that preserves this property.

\begin{figure}
  \begin{center}
    \widfig{0.4}{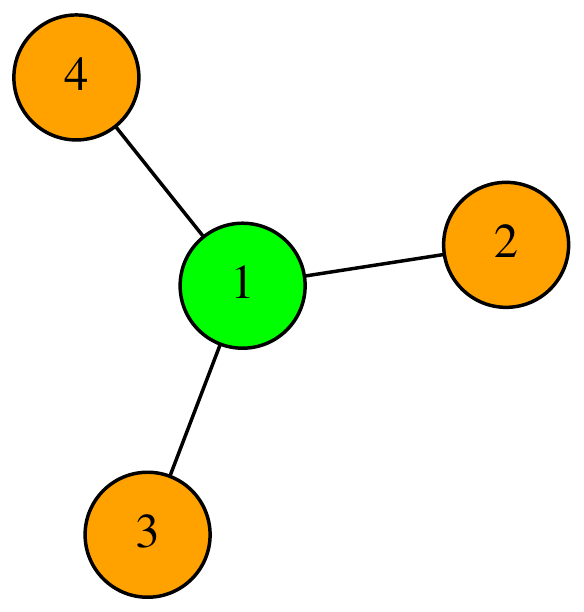}
  \end{center}
  \caption{\label{fig:star4}(color online) The star-like network,
    $N=4$.}
\end{figure}

We consider star-like networks, see Fig.~\ref{fig:star4}. Let
$\meye{m}$ be a unit $m\times m$ matrix, $\mzergen{m}{n}$ be a
$m\times n$ matrix of zeros, and $\monegen{m}{n}$ be a $m\times n$
matrix of ones. Using this notation the adjacency matrix can written
in block form as
\begin{equation}
  \label{eq:adj_gen}
  \mat A=
  \begin{pmatrix}
    \mzer{1} & \vtone{M} \\
    \vone{M} & \mzer{M}
  \end{pmatrix}
\end{equation}
where $M=N-1$.

In numerical simulations some regimes was found to emerge outside the
theoretical ranges of stability, computed in
Sec.~\ref{sec:stab-cert-regim}. In particular this is the case for
full chaotic synchronization. This spurious stability occurs due to
round-off errors in numerical model. To eliminate it a very weak noise
is added to system variables:
\begin{equation}
  \label{eq:small_noise}
  x_n(t)\to x_n(t)+\xi(2t),\;y_n(t)\to y_n(t)+\beta \xi(2t+1),
\end{equation}
where $\xi(t)\in(-10^{-12},10^{-12})$. Notice that the noise is added
only to fix the drawback of the numerical model. Since its amplitude
is not much higher then the machine epsilon for employed variables,
$\approx 10^{-16}$, it is indistinguishable in the resulting time
series. Thus, we emphasize that the study of noise influence on the
network lies outside the scope of our paper.

\section{Criteria of synchronization}\label{sec:detect-synchr}

Coincidence of local extrema of two discrete time signals can be
treated as their phase synchronization while the relative frequency of
the coincidences quantifies the degree of the
synchronization~\cite{JalanAmritkar2003,JalanAmritkar2005}. Below we
consider a generalization of this approach taking into account both
dip-dip and peak-dip coincidences.

Let us consider the nodes $m$ and $n$, $m<n$. Given a starting time
$t_0$ and the interval $\clutim$, count at $t_0\leq t<t_0+\clutim$ the
numbers $\nu_m$ and $\nu_n$ of local minima of $x_m(t)$ and $x_n(t)$,
respectively, and the number $\nu_{mn}$ of simultaneous minima of
$x_m$ and $x_n$. In addition we count the number of times $\mu_{mn}$
when the local minima of $x_m$ occur simultaneously with the local
maxima of $x_n$, and the number of times $\mu_{nm}$ when the minima of
$x_n$ coincide with the maxima of $x_m$. Note that since $m<n$,
$\mu_{mn}$ can be treated as elements of the upper triangle of a
square matrix and $\mu_{nm}$ form its lower triangle. Then the dip-dip
distance can be computed as
\begin{equation}
  \label{eq:inphase_dist}
  \ddip_{m\dash n}=
  \begin{cases}
    1-\nu_{mn}/\max\{\nu_m,\nu_n\} & \text{if } \max\{\nu_m,\nu_n\}>0, \\
    0 & \text{if } \max\{\nu_m,\nu_n\}=0.
  \end{cases}
\end{equation}
The second choice covers the situation when both $x_m$ and $x_n$ do
not oscillate at all. This value is introduced in
Refs.~\cite{JalanAmritkar2003,JalanAmritkar2005} as phase
distance. Analogously we can define peak-dip distance as
\begin{equation}
  \label{eq:contphase_dist}
  \dpik_{m\dash n}=
  \begin{cases}
    1-\mu_{mn}/\max\{\nu_m,\nu_n\} & \text{if } \max\{\nu_m,\nu_n\}>0, \\
    0 & \text{if } \max\{\nu_m,\nu_n\}=0.
  \end{cases}
\end{equation}
Two nodes are did-dip or peak-dip synchronized on the interval
$\clutim$ if $\ddip_{m\dash n}=0$ or $\dpik_{m\dash n}=0$,
respectively.

The full synchronization can be detected using $\dful_{m\dash n}$:
\begin{equation}
  \label{eq:f_inphas_dist}
  \dful_{m\dash n}=\variance\{x_m(t)-x_n(t)\;|\;t_0\leq t<t_0+\clutim\},
\end{equation}
where $\variance$ stands for variance. Moreover we are going to detect
complementary synchronization via the vanish of $\dcom_{m\dash n}$:
\begin{equation}
  \label{eq:f_contphas_dist}
  \dcom_{m\dash n}=\variance\{x_m(t)+x_n(t)\;|\;t_0\leq t<t_0+\clutim\}.
\end{equation}
The synchronization is actually registered when $\dful_{m\dash n}$ or
$\dcom_{m\dash n}$ are below the threshold $10^{-12}$, which the level
of the added noise, see Eq.~\eqref{eq:small_noise}.

All of the above criteria depend on $\clutim$. It determines the
resolution of the detection procedure. If $\clutim\to\infty$ the
procedure responds only to full-time regimes. So it is preferable for
$\clutim$ to be as short as possible. In this case in addition to
full-time synchronization, one can detect synchronization windows as
series of subsequent intervals where $\ddip_{m\dash n}$,
$\dpik_{m\dash n}$, $\dful_{m\dash n}$, or $\dcom_{m\dash n}$ vanish.
The lower boundary for $\clutim$ is the average interval between peaks
and dips: $\clutim$ has to be large enough to catch at least two-three
of them. In simulations below we set $\clutim=16$.

\section{Classification of dynamical
  regimes}\label{sec:dynamics-star-like}

\subsection{Method of analysis}

Since it is unclear \emph{a priori} what types of behaviour can be
observed, the most reliable way is the visual inspection of time
series corroborated by some appropriate characteristic numbers.

First of all we are going to employ the first Lyapunov exponent whose
positive value indicates chaos, the negative sign reveals periodicity
and the zero means quasi-periodicity\footnote{In principle, a negative
  Lyapunov exponent can also be observed for so called strange
  non-chaotic attractors~\cite{SNA}. However, we did not encountered
  them in our case.}. One has to take many random initial conditions
and compute $\lambda_1$ for each corresponding trajectory. The
resulting values are grouped very well near a few points representing
different regimes. This approach is usually used for analysis of
multistability~\cite{Pisarchik2014167}.

Various types of synchronous oscillations will be detected using
criteria introduced in Sec.~\ref{sec:detect-synchr}. We take
$\clutim=16$ and test at each subsequent interval $\clutim$ if one of
the four characteristic values vanishes for each pair of
oscillators. If yes, we are inside a window of synchronization of a
corresponding type. Then we find the largest lengths of the windows
that are registered along the observation time, which is
$t_{\text{max}}=10^{5}\clutim$. The full-time synchronization is
registered if the corresponding window lasts during the whole
observation time $t_{\text{max}}$.

\subsection{The star with $N=4$}\label{sec:star-n4}

\begin{figure*}
  \centering
  \fulfig{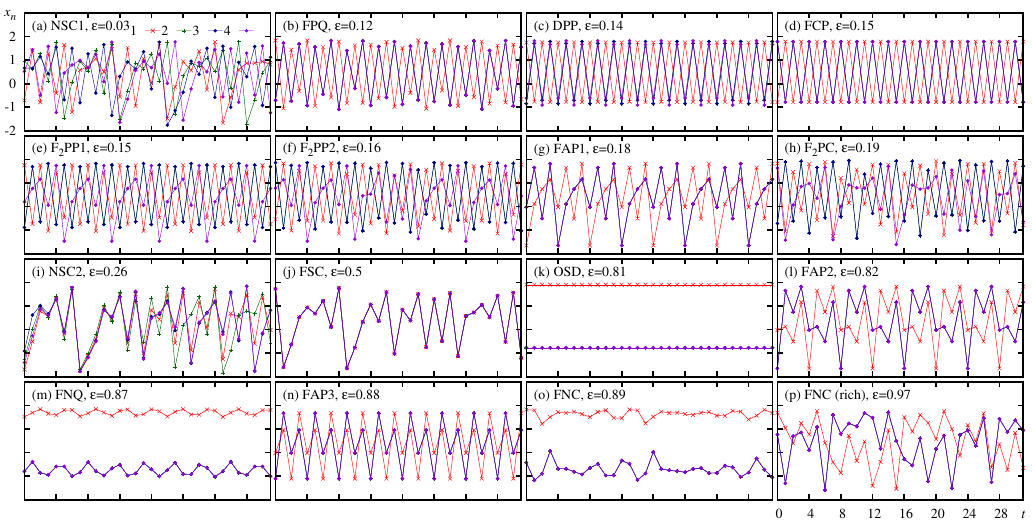}
  \caption{\label{fig:skr}(color online) Various regimes of the
    star-like network \eqref{eq:netw_henon},~\eqref{eq:adj_gen} with
    $N=4$. See details in the text.}
\end{figure*}

\begin{figure}
  \centering
  \widfig{1}{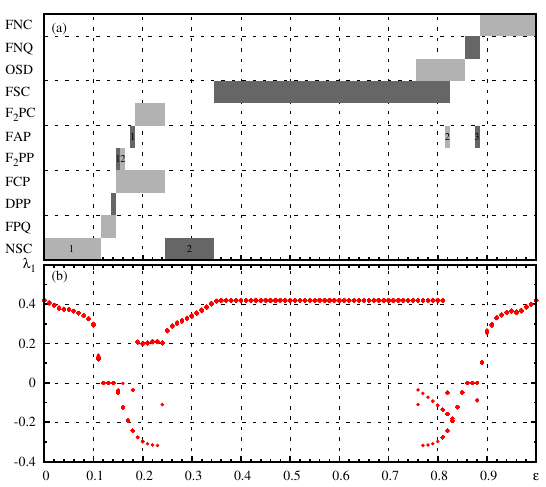}
  \caption{\label{fig:rgline}(color online) (a) An ordering of the
    regimes shown in Fig.~\ref{fig:skr}. See also
    Tab.~\ref{tab:regimes}. (b) The first Lyapunov exponents computed
    for 200 random initial conditions at each
    $\epsilon=0,0.01,0.02,\ldots,1$}
  \medskip
\end{figure}

Consider dynamics of the network~\eqref{eq:netw_henon},
\eqref{eq:adj_gen} when the coupling $\epsilon$ varies from 0 to 1. We
have inspected all values of $\epsilon\in[0,1]$ with the step $0.01$
for the smallest nontrivial star $N=4$. Moreover when a regime of some
sort appeared only on a single step, we also checked if it existed at
least in a small vicinity of the corresponding $\epsilon$ to make sure
that this is a typical situation. To detect a multistability, for each
$\epsilon$ we tested at least 200 random initial conditions. The
results are gathered in Figs.~\ref{fig:skr} and \ref{fig:rgline} that
are discussed below.

To facilitate the referencing we will label the regimes with three
letters. The first two ones are an abbreviation describing a type of
synchronization between network nodes. In some cases we will supply
them with a subscript to show a number of involved nodes. The third
letter indicates the character of oscillations: periodic ``P'',
quasiperiodic ``Q'', chaotic ``C''. If there are several regimes with
identical abbreviations we also add an index number. The exception to
this scheme is the oscillation death that will be labeled merely as
OSD.

\begin{table}
  \caption{\label{tab:regimes} Stability ranges of $\epsilon$ for
    regimes in Figs.~\ref{fig:skr} and \ref{fig:rgline} obtained via
    step by step analysis of dynamics with $\Delta\epsilon=0.01$. 
    More accurate values for FSC, 
    OSD and FCP are found in Sec.~\ref{sec:stab-cert-regim}, see
    Eqs.~\eqref{eq:range_fsc},~\eqref{eq:range_osd}, and
    \eqref{eq:range_fsp}, respectively.}
  \begin{tabular}{rlrl}
    \hline
    \hline
    NSC: & [0,0.11], [0.25,0.34] & FPQ: & [0.12,0.14] \\
    DPP: & 0.14 & FCP: & [0.15,0.24] \\
    F\subtwo PP: & 0.15, 0.16  & FAP: & 0.18, 0.82, 0.88 \\
    F\subtwo PC: & [0.19,0.24] & FSC: & [0.35,0.82] \\
    OSD: & [0.76,0.85] & FNQ: & [0.86,0.88] \\
    FNC: & [0.89,1] & {} & {} \\
    \hline
    \hline
  \end{tabular}
\end{table}

The ordering of the regimes is shown in Fig.~\ref{fig:rgline}(a), and
the boundaries of the regimes are collected in
Tab.~\ref{tab:regimes}. Fig.~\ref{fig:rgline}(b) shows $\lambda_1$
vs. $\epsilon$ computed for 200 random initial conditions.

\begin{figure}
  \centering
  \widfig{1}{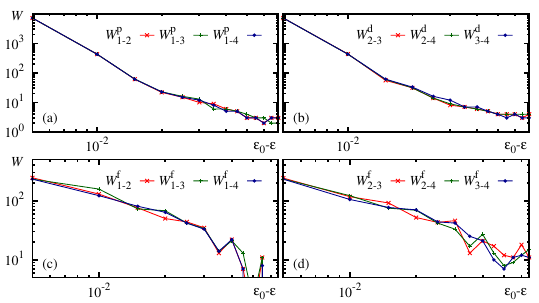}
  \caption{\label{fig:win_nsc}(color online) Longest synchronization
    windows vs. coupling near the boundaries of ranges. (a,b) NSC1,
    $\epsilon_0=0.115$; (c,d) NSC2,
    $\epsilon_0=0.345$. $W^{\,\text{d}}_{m\dash n}$,
    $W^{\,\text{p}}_{m\dash n}$, and $W^{\,\text{f}}_{m\dash n}$
    denote the lengths of the longest windows where corresponding
    $\ddip_{m\dash n}$, $\dpik_{m\dash n}$, and $\dful_{m\dash n}$
    vanish.}
\end{figure}

\neweps{$\epsilon=0,0.01,\ldots,0.11$} \textbf{NSC1}\footnote{No
  Synchronization, Chaos}. There is only one chaotic attractor here.
This is confirmed by the coincidence in Fig.~\ref{fig:rgline}(b) of
Lyapunov exponents, and by visual inspection of time series whose
typical example is shown in Fig.~\ref{fig:skr}(a). The power law
divergence of peak-dip synchronization windows between the hub and
subordinates and dip-dip windows for the subordinates, see
Figs.~\ref{fig:win_nsc}(a) and (b), indicate the intermittency
appearing near the right boundary of the considered range.

\neweps{$\epsilon=0.12,0.13$} \textbf{FPQ}\footnote{Full synchronization
  of the subordinates, $\dful_\subsub=0$, and Peak-dip synchronization
  of the hub with them, $\dpik_\hubsub=0$, Quasiperiodicity},
Fig.~\ref{fig:skr}(b). The quasiperiodicity is confirmed by the vanish
of $\lambda_1$ in Fig.~\ref{fig:rgline}(b). 

\neweps{$\epsilon=0.14$} \textbf{FPQ} and \textbf{DPP}\footnote{Dip-dip
  synchronization of the subordinates, $\ddip_\subsub=0$, and Peak-dip
  synchronization of the hub with them, $\dpik_\hubsub=0$,
  Periodicity}, Fig.~\ref{fig:skr}(c). Actually in DPP two
subordinates are fully synchronized while the third one, the node
number 4, almost coincides with them. The Lyapunov exponent
corresponding to DPP is close to zero, $\lambda_1=-0.00030$, and one
can not distinguish it in Fig.~\ref{fig:rgline}(b). However, the
inspection of data reveals that it is strictly smaller then the
Lyapunov exponent for FPQ, so that despite of FPQ this regime is
periodic. Moreover, we computed Fourier spectra (not shown), that
again confirmed the quasiperiodicity of FPQ and the periodicity of
DPP.

\neweps{$\epsilon=0.15$} \textbf{FCP}\footnote{Full synchronization of
  the subordinates, $\dful_\subsub=0$, and Complementary
  synchronization of the hub with them, $\dcom_\hubsub=0$,
  Periodicity} and \textbf{F\subtwo PP1}\footnote{Full synchronization
  of two of the subordinates, $\dful_{2\dash 3}=0$, Peak-dip
  synchronization of the hub with this two, $\dpik_{1\dash 2,3}=0$,
  the third subordinate oscillates separately, Periodicity}. See
Figs.~\ref{fig:skr}(d) and (e), respectively. The dominating regime
here is FCP. It exists within a wide range of $\epsilon$, see
Tab~\ref{tab:regimes}. In Fig.~\ref{fig:rgline}(b) it corresponds to
the lowest branch of the Lyapunov exponents. The F\subtwo PP1 has
smaller basing of attraction and exists within a narrow range of
$\epsilon$ around $0.15$. At $\epsilon=0.15$ the Lyapunov exponents
for these two regimes though different, are close to each other and
barely distinguishable in Fig.~\ref{fig:rgline}(b): for FCP
$\lambda_1=-0.0487$ and for F\subtwo PP1 $\lambda_1=-0.0389$.

\neweps{$\epsilon=0.16$} \textbf{FCP} and \textbf{F\subtwo PP2}. Here
we have the second version of the periodic regime with full
synchronization of two subordinates, see Fig.~\ref{fig:skr}(f). Though
it looks like F\subtwo PP1, the closer inspection reveals that the
period of F\subtwo PP1 is 6 and the period of F\subtwo PP2 is 20. The
first Lyapunov exponent for F\subtwo PP2 is $\lambda_1=-0.00191$.

\neweps{$\epsilon=0.17$} \textbf{FCP}. 

\neweps{$\epsilon=0.18$} \textbf{FCP} and \textbf{FAP1}\footnote{Full
  synchronization of the subordinates, $\dful_\subsub=0$, and
  Anti-phase synchronization of the hub with them, Periodicity},
Fig.~\ref{fig:skr}(g). The Lyapunov exponents are $\lambda_1=-0.243$,
and $\lambda_1=-0.0360$, respectively.

\neweps{$\epsilon=0.19,\ldots,0.24$} \textbf{FCP} and \textbf{F\subtwo
  PC}\footnote{All as for F\subtwo PP1, but chaos instead of
  periodicity}, Fig.~\ref{fig:skr}(h). The emergence of the second
regime is clearly seen in Fig.~\ref{fig:rgline}(b), where the upper
branch of positive Lyapunov exponents appears near $0.2$. Note that
$\lambda_1$ for F\subtwo PC is rather independent on $\epsilon$.

\neweps{$\epsilon=0.25,\ldots,0.34$} \textbf{NSC2},
Fig.~\ref{fig:skr}(i). In the second version of these regime we again
observe an intermittency.  When $\epsilon$ approaches the right
boundary of the range the full synchronization windows demonstrate
power law divergence, see Fig.~\ref{fig:win_nsc}(c,d).

\neweps{$\epsilon=0.35,\ldots,0.75$} \textbf{FSC}\footnote{Full
  Synchronization, Chaos}, Fig.~\ref{fig:skr}(j). In
Fig.~\ref{fig:rgline}(b) this regime is represented by a perfect
horizontal line.

\neweps{$\epsilon=0.76,\ldots,0.81$} \textbf{FSC} and
\textbf{OSD}\footnote{Oscillation Death}, Fig.~\ref{fig:skr}(k). OSD
has very small basin of attraction here. It becomes more or less
easily detectable only at $\epsilon=0.81$. There are two forms of
OSD. The second one, not shown, is an interchange of the values for
the hub and subordinate node variables. Two forms have different
negative Lyapunov exponents, so that there are three branches in
Fig.~\ref{fig:rgline}(b), one positive for FSC and two negative for
OSD.

\neweps{$\epsilon=0.82$} \textbf{FSC}, \textbf{OSD}, and
\textbf{FAP2}, Fig.~\ref{fig:skr}(l). Accordingly, there are four
values of Lyapunov exponents, see Fig.~\ref{fig:rgline}(b):
$\lambda_1=-0.243$ and $-0.158$ for OSD, $\lambda_1=-0.0492$ for FAP2,
and $\lambda_1=0.419$ for FSC.

\neweps{$\epsilon=0.83$} \textbf{OSD}. The Lyapunov exponents
corresponding to the two forms of OSD becomes very close to each
other, $\lambda_1=-0.192$ and $\lambda_1=-0.182$. Right after this
point they merge, see Fig.~\ref{fig:rgline}(b).

\neweps{$\epsilon=0.84,0.85$} \textbf{OSD}. Now both forms of this
regime have identical Lyapunov exponents, see
Fig.~\ref{fig:rgline}(b).

\neweps{$\epsilon=0.86,87$} \textbf{FNQ}\footnote{Full synchronization
  of the subordinates, $\dful_\subsub=0$, and No synchronization of
  the nub with them, Quasiperiodicity}. This regime appears when two
fixed points corresponding to OSD become unstable, see
Fig.~\ref{fig:skr}(m). The corresponding Lyapunov exponent is zero
here, see Fig.~\ref{fig:rgline}(b).

\neweps{$\epsilon=0.88$} \textbf{FNQ}, \textbf{FAP3},
Fig.~\ref{fig:skr}(n). The Lyapunov exponents are zero for FNQ and
$\lambda_1=-0.0888$ for FAP3, see Fig.~\ref{fig:rgline}(b).

\neweps{$\epsilon=0.89,\ldots,1$} \textbf{FNC}\footnote{Same as FNQ,
  but chaos instead of quasiperiodicity}. Starting from this point the
oscillations born from OSD fixed points become chaotic. Initially the
oscillations have sufficiently small amplitude, see
Fig.~\ref{fig:skr}(o), and as $\epsilon$ grows the oscillations become
more and more entangled, see Fig.~\ref{fig:skr}(p).

\subsection{Stars with $N>4$ and wild multistability}
  
\begin{figure}
  \centering
  \widfig{1}{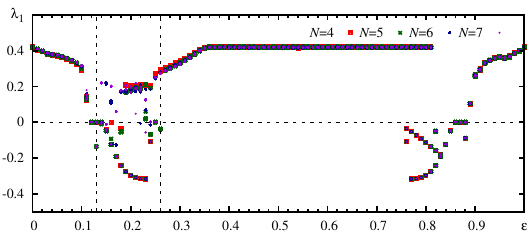}
  \caption{\label{fig:starlyeps}(color online) The fist Lyapunov
    exponents for $N=4,5,6$ and $7$. Data for $N=4$ are the same as in
    Fig.~\ref{fig:rgline}(b). Vertical lines at $\epsilon=0.13$ and
    $0.26$ mark an area of wild multistability where the Lyapunov
    exponents for different $N$ do not coincide.}
\end{figure}

Figure~\ref{fig:starlyeps} compares first Lyapunov exponents computed
for stars of different sizes. Observe remarkable coincidence of almost
all points, indicating the identity of dynamics regardless of
$N$. Additionally we verified it by visual inspection of time series
and by the analysis of synchronization windows. Moreover, below we
will provide rigorous proofs for certain regimes that their ranges of
stability do not depend on $N$.

However, one have also to notice the area where $\lambda_1$ does
depend on $N$. This area is marked in Fig.~\ref{fig:starlyeps} by
dashed vertical lines. Some of attractors like FPQ ($\lambda_1=0$) and
FCP (the lowest branch of $\lambda_1$) still exists regardless of
$N$. But in addition a variety of other attractors appears only at
certain $N$, each with a narrow range of stability on $\epsilon$.

\begin{figure}
  \centering
  \widfig{1}{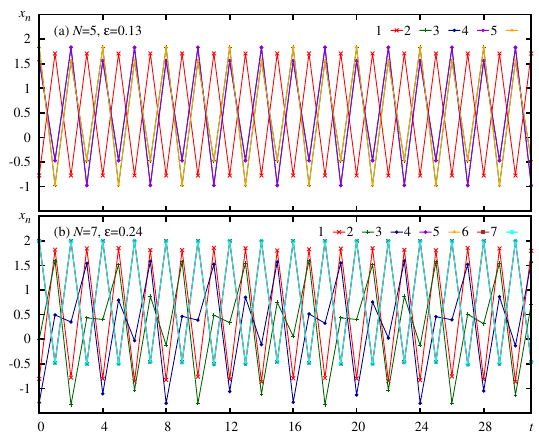}
  \caption{\label{fig:skrbig}(color online) Examples of dynamics of
    the star-like networks with $N>4$ within the range of wild
    multistability. (a) $N=5$, $\epsilon=0.13$ (b) $N=7$,
    $\epsilon=0.24$}
\end{figure}

Figure~\ref{fig:skrbig} illustrates what can be encountered within
this area.  Figure~\ref{fig:skrbig}(a) demonstrates a periodic regime
with $\lambda_1=-0.136$, emerging at $N=5$, and $\epsilon=0.13$. Here
the subordinates are fully synchronized by pairs:
$\dful_{2\dash 5}=0$, and $\dful_{3\dash 4}=0$.  These pairs in turn
are dip-dip synchronized with each other, $\ddip_{2,3\dash 4,5}=0$.
The hub is in peak-dip synchronization with all subordinates,
$\dpik_{1\dash 2,3,4,5}=0$.  Figure~\ref{fig:skrbig}(b) represents a
chaotic regime at $N=7$ and $\epsilon=0.24$ with
$\lambda_1=0.00206$. Four of six subordinates, namely 4,5,6, and 7,
are fully synchronized. Two others, 2 and 3, are dip-dip synchronized
with each other. This pair is in turn in peak-dip synchronization with
the other four. The hub is dip-dip synchronized with the nodes 2 and 3
and peak-dip synchronized with the other four.

Altogether, within the area delimited by the dashed lines in
Fig.~\ref{fig:starlyeps} there is the large number of attractors with
highly interwoven basins and narrow stability ranges. As a result one
can hardly predict the qualitative nature of the expected dynamics
since each perturbation to $\epsilon$ or initial condition results in
quite different behavior. Similar situation was already reported for
systems with high dimensional phase space, in particular, for coupled
map lattices, see the review in Ref.~\cite{Feudel2008}. We suggest to
refer to it as \emph{wild multistability} to emphasize its amazing
richness.

One more similar area is located in Fig.~\ref{fig:starlyeps} between
$\epsilon\approx 0.8$ and $\epsilon\approx 0.9$. Though the Lyapunov
exponents do not depend on $N$ here, the basins of coexisting
attractors can be colocated in complicated manner impeding prediction
of the dynamics. 

\subsection{Remote synchronization}

As we already mentioned above, remote synchronization occurs between
two or more nodes without direct connections, but linked with the
common hub node. The important point here is that the hub remains not
synchronized with them, that is achieved by detuning its natural
frequency~\cite{RemSyn1,RemSyn2}.

In our case there are regimes with fully synchronous subordinates and
the hub not coinciding with them. These are FPQ, DPP, FCP, all
F\subtwo PPs, all FAPs, and F\subtwo PC, see Fig.~\ref{fig:skr}. This
behavior though reminds remote synchronization, can not be classified
like this since actually the hub is also synchronized with the rest of
nodes with a phase shift.

The remote synchronization is nevertheless observed for our system,
see FNQ and FNC in Figs.~\ref{fig:skr}(m) and (o,p),
respectively. These regimes essentially differ from those reported in
Ref.~\cite{RemSyn1}.  First, the subordinates are fully synchronized,
while in Ref.~\cite{RemSyn1} phase synchronization is
reported. Second, in our case oscillators are identical, but the hub
still does not get synchronized with the subordinates. Third, FNQ and
FNC are quasiperiodic and chaotic regimes, respectively, contrary to
the periodic case reported in~\cite{RemSyn1}. For these regimes the
difference between the hub and subordinates is found to be not
necessary to prevent their synchronization.

\section{Stability of certain regimes}\label{sec:stab-cert-regim}

\subsection{Full chaotic synchronization (FSC)}\label{sec:full-chaot-synchr}

Stability of the fully synchronized state can be analyzed using so
called Master Stability Function
(MSF)~\cite{MSF98,Boccaletti2006175,Pecora98}. First we need the Jacobian
matrix of the network~\eqref{eq:netw_henon}. It has a block form being
composed of $N\times N$ matrices:
\begin{equation}
  \label{eq:henon_jac}
  \mat J(t)=
  \begin{pmatrix}
    \mat F(t) & \identmat \\
    \beta\identmat & 0
  \end{pmatrix},
\end{equation}
where 
\begin{equation}
  \label{eq:jac_matr}
  \begin{gathered}
    \mat F(t)=-2\mat G(t)\, [(1-\epsilon)\identmat+\epsilon 
    \mat K^{-1}\mat A],\\
    \mat G(t)=\diagmat\{x_n+\epsilon h_n\},\; \mat K=\diagmat\{k_n\},
  \end{gathered}
\end{equation}
and $\identmat$ is the identity matrix~\cite{NWL2014}.

\begin{figure}
  \centering
  \widfig{1}{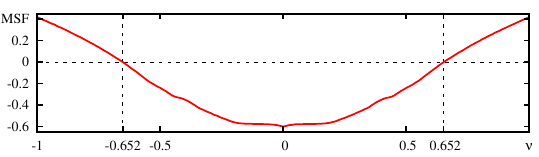}
  \caption{\label{fig:msf}(color online) MSF for the
    network~\eqref{eq:netw_henon}, \eqref{eq:adj_gen}. Vertical dotted
    lines delimit an area where MSF is negative.}
\end{figure}

Jacobian matrix at $x_n(t)\to x(t)$ governs the evolution of tangent
perturbation to the synchronization manifold, where $x(t)$ is produced
by a local isolated oscillator. Decomposing the perturbations over
eigenvectors of the matrix $(\mat K^{-1}\mat A)$ one obtains the
following equations for perturbation amplitudes:
\begin{equation}
  \label{eq:msf_eqn}
  \begin{aligned}
    \delta p(t+1)&=-\nu[2x(t)\delta p(t)]+\delta q(t),\\
    \delta q(t+1)&=\beta \, \delta p,
  \end{aligned}
\end{equation}
where 
\begin{equation}
  \label{eq:msf_munu}
  \nu=1+\epsilon\,(\phi-1),
\end{equation}
and $\phi$ is an eigenvalue of $(\mat K^{-1}\mat A)$. We can treat
$\nu$ a free parameter and define MSF as a conditional Lyapunov
exponent, i.e., an average rate of exponential growth of a solution of
Eq.~\eqref{eq:msf_eqn} when $x(t)$ runs along a trajectory of the
local oscillator. The graph of MSF is shown in
Fig.~\ref{fig:msf}. Note that MSF is negative at $|\nu|<0.652$.

Using Eq.~\eqref{eq:adj_gen}, one can rewrite matrix
$(\mat K^{-1}\mat A)$ as
\begin{equation}
  \label{eq:kadj}
  \mat K^{-1}\mat A=
  \begin{pmatrix}
    \mzer{1} & \vtone{M}/M \\
    \vone{M} & \mzer{M}
  \end{pmatrix},
\end{equation}
and then compute its eigenvalues:
\begin{equation}
  \label{eq:adj_eival}
  \phi_1=1,\phi_2=-1,\phi_3=\phi_4=\cdots=\phi_N=0
\end{equation}
The eigenvector corresponding to $\phi_1$ has identical elements being
responsible for chaotic dynamics on the synchronization manifold. All
other eigenvectors and eigenvalues correspond to transverse
perturbations whose vanish is the necessary condition of stability of
the full synchronization. 

Substituting $\phi=-1$ and $\phi=0$ to Eq.~\eqref{eq:msf_munu}, one
can see that MSF is negative for all transverse perturbations when
\begin{equation}
  \label{eq:range_fsc}
  \epsilon=\epsilon_{\text{FSC}}\in[0.348,0.826].
\end{equation}
Withing this range the full synchronization attractor is transversely
stable on average. For attractors with regular structure this is also
a sufficient condition, but when the dynamics is chaotic, the
synchronization can be destabilized by a small noise even when MSF is
negative. This is due to the presence of transversally unstable
invariant sets (cycles, in particular) embedded into the
synchronization manifold, see~\cite{milnor2004concept}. In our case
however the influence of noise~\eqref{eq:small_noise} is
indistinguishably weak. One can see in Fig.~\ref{fig:rgline}(b) that
the first Lyapunov exponent attains the level corresponding to this
regime almost exactly within the range~\eqref{eq:range_fsc}.

Note that the stability range for the full chaotic synchronization
does not depend on $N$. The reason is that the matrix
$(\mat K^{-1}\mat A)$ has only three different eigenvalues $1$,$-1$,
and $0$ regardless of $N$.

\subsection{Oscillations death (OSD)}\label{sec:oscill-death-osd}

Let $x_a$ and $x_b$ be states of the hub and the subordinates at OSD,
respectively. Using Eqs.~\eqref{eq:netw_henon}, one can write:
\begin{equation}
  \begin{aligned}
    x_a&=\alpha-[x_a+\epsilon(x_b-x_a)]^2+\beta x_a,\\
    x_b&=\alpha-[x_b+\epsilon(x_a-x_b)]^2+\beta x_b.
  \end{aligned}
\end{equation}
The solution of these equations can be expressed via roots $\xi_1$ and
$\xi_2$ of the polynomial
\begin{equation}
  \label{eq:osd_state}
  \xi^2+\frac{\beta-1}{2\epsilon-1}\xi+
  \frac{((\beta-1)^2-4\alpha)\epsilon^2+\alpha(4\epsilon-1)}
  {(2\epsilon-1)^4}=0.
\end{equation}
There are two couples, $x_a=\xi_1$, $x_b=\xi_2$, and $x_a=\xi_2$,
$x_b=\xi_1$, corresponding to two forms of OSD.

Stability of the OSD solution is determined by eigenvalues $\mu$ of
the Jacobian matrix~\eqref{eq:henon_jac} at ($x_a$, $x_b$). The
eigenvalue problem for $\mu$ is reduced to
\begin{equation}
  \label{eq:jac_osd_eival}
  \mu^2-\mu f-\beta=0,
\end{equation}
where $f$ is an eigenvalue of $\mat F$, see
Eq.~\eqref{eq:jac_matr}. One can see that each $f$ produces a couple
of $\mu$. Computing $f$ we employ the block representation of $\mat A$
\eqref{eq:adj_gen}:
\begin{equation}
  \label{eq:matf_osd}
  \mat F=
  \begin{pmatrix}
    \kappa_a (1-\epsilon)\meye{1} & (\kappa_a \epsilon/M) \vtone{M}\\
    \kappa_b \epsilon \vone{M} & \kappa_b (1-\epsilon) \meye{M}
  \end{pmatrix}.
\end{equation}
Here $\kappa_a=x_a+\epsilon(x_b-x_a)$ corresponds to the hub node and
$\kappa_b=x_b+\epsilon(x_a-x_b)$ comes from the subordinates. The
eigenproblem for $\mat F$ results in
\begin{equation}
  \label{eq:feival_osd}
  [f+2\kappa_a(1-\epsilon)][f+2\kappa_b(1-\epsilon)]M=
  4\kappa_a\kappa_b\epsilon^2\delta.
\end{equation}
Here $\delta$ is one of eigenvalues of $\mone{M}$,
\begin{equation}
  \label{eq:mone_eival}
  \delta_1=M,\delta_2=\delta_3=\cdots=\delta_M=0.
\end{equation}
Each $\delta=0$ produces a couple of solutions of
Eq.~\eqref{eq:feival_osd}. However, $f=-2\kappa_a(1-\epsilon)$ is
spurious that can be verified by substituting it to the eigenproblem
for $\mat F$. Thus $\mat F$ has the eigenvalue
$f=-2\kappa_b(1-\epsilon)$ with the multiplicity $M-1$. Two more
eigenvalues correspond to $\delta=M$. They can be found from
Eq.~\eqref{eq:feival_osd} by canceling $M$. Note that none of $f$
depends on the number of the network nodes.

\begin{figure}
  \centering
  \widfig{1}{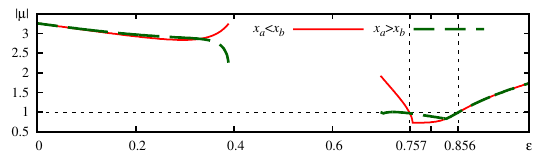}
  \caption{\label{fig:osd_stab}(color online) The largest by magnitude
    eigenvalue of the Jacobian matrix~\eqref{eq:henon_jac},
    \eqref{eq:jac_matr} for OSD. Vertical lines delimit the range of
    stability.}
\end{figure}

To obtain the stability range for OSD one has to vary $\epsilon$,
compute $f$, and find the absolute values of corresponding eigenvalues
$\mu$ from Eq.~\eqref{eq:jac_osd_eival}. The regime is stable when
$|\mu|<1$. Figure~\ref{fig:osd_stab} shows the plots of the largest by
magnitude $|\mu|$. In the middle area the curves vanish because
Eq.~\eqref{eq:osd_state} has complex roots. Thus the range of
stability computed in this way is
\begin{equation}
  \label{eq:range_osd}
  \epsilon=\epsilon_{\text{OSD}}\in[0.757,0.856].
\end{equation}
It agrees well with the range obtained via straightforward
simulations, see Tab.~\ref{tab:regimes}.

\subsection{Full synchronization of the subordinates, and
  complementary synchronization of the hub with them
  (FCP)}\label{sec:fcp-range}

In this regime the system oscillates with period two so that the hub
and the coinciding subordinates switch between $x_a$ and $x_b$:
\begin{equation}
  \begin{aligned}
    x_a&=\alpha-[x_b+\epsilon(x_a-x_b)]^2+\beta x_a,\\
    x_b&=\alpha-[x_a+\epsilon(x_b-x_a)]^2+\beta x_b.
  \end{aligned}
\end{equation}
Solutions to this equation set can be found as roots of the polynomial
\begin{equation}
  \label{eq:fcp_state}
  \xi^2-\frac{\beta-1}{2\epsilon-1}\xi+
  \frac{(1-\epsilon)[(1-\beta)^2(1-\epsilon)+4\alpha\epsilon]-\alpha}
  {(2\epsilon-1)^4}=0.
\end{equation}

Stability of FCP is determined by the eigenvalues $\mu$ of the matrix
$(\mat J_b\mat J_a)$ where $\mat J_a$ is computed as in
Sec.~\ref{sec:oscill-death-osd}, and $\mat J_b$ is similar on
interchanging $x_a$ and $x_b$:
\begin{equation}
  (\mu-\beta)^2=f\mu,
\end{equation}
where $f$ is an eigenvalue of the matrix product $(\mat F_b\mat F_a)$.

Using Eq.~\eqref{eq:matf_osd} one can write the matrix
$(\mat F_b\mat F_a)$ explicitly and find that its eigenproblem results
in the equation
\begin{equation}
  \label{eq:feival_fcp}
  (f-a_{11})(f-b_{22})M=(a_{22}f-a_{11}a_{22}+a_{12}a_{21})\delta,
\end{equation}
where $a_{11}=\kappa_b[\kappa_a(1-\epsilon)^2+\kappa_b\epsilon^2]$,
$a_{12}=\kappa_b(\kappa_a+\kappa_b)\epsilon(1-\epsilon)$,
$a_{21}=\kappa_a(\kappa_a+\kappa_b)\epsilon(1-\epsilon)$,
$a_{22}=\kappa_a^2\epsilon^2$,
$b_{22}=\kappa_a\kappa_b(1-\epsilon)^2$, and $\delta$ is an eigenvalue
of $\mone{M}$, see Eq.~\eqref{eq:mone_eival}.  The root $f=a_{11}$ at
$\delta=0$ is spurious that can be tested by a direct
substitution. Thus there is an eigenvalue $f=b_{22}$ with the
multiplicity $M-1$ and two more eigenvalues are the solutions of
Eq.~\eqref{eq:feival_fcp} at $\delta=M$. Notice, that as in
Sec.~\ref{sec:oscill-death-osd}, $M$ is canceled so that the stability
range again does depend of the network size.

\begin{figure}
  \centering
  \widfig{1}{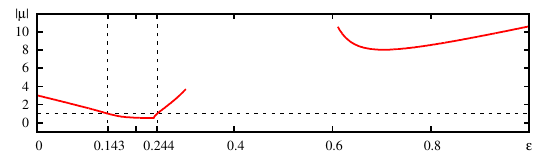}
  \caption{\label{fig:fcp_stab}(color online) The largest by
    magnitude eigenvalue of the Jacobian matrix for the regime
    FCP. Vertical lines delimit the range of stability.}
\end{figure}

Varying $\epsilon$ we can compute $f$ and then find the largest
$|\mu|$. The plot is shown in Fig.~\ref{fig:fcp_stab}. The FCP regime
is stable where $|\mu|<1$. This range is
\begin{equation}
  \label{eq:range_fsp}
  \epsilon=\epsilon_{\text{FCP}}\in[0.143,0.244].
\end{equation}

\section{Outline and conclusions}

We considered a variety of dynamics of H\'{e}non map networks with
star-like topology. This is found to be amazingly rich. In brief, as
the coupling strength grows from zero to one the following areas are
observed.
\begin{itemize}
\item Non-synchronized oscillations, intermittency of synchronization
  windows. 
\item Quasiperiodicity.
\item Wild multistability. Small variations of initial conditions
  result in periodic, quasiperiodic and chaotic solutions. The zoo of
  regimes is very sensitive both to variations of the coupling
  strength and to the network size.
\item Again non-synchronized oscillations, intermittency of synchronization
  windows.
\item Full chaotic synchronization of all network nodes.
\item Coexistence of the full chaotic synchronization, oscillation
  death and periodic oscillations when the hub oscillates in
anti-phase with fully synchronized subordinates.
\item Oscillation death.
\item Remote synchronization of quasiperiodic oscillations. At narrow
  range of coupling values it coexists with another version of the
  regime of periodic oscillations when the hub is in anti-phase with
  fully synchronized subordinates.
\item Remote synchronization of chaotic oscillations.
\end{itemize}

We conjecture that the list of regimes and corresponding stability
ranges in most cases remain the same regardless of $N$. This is
checked numerically for $N\leq 7$ and rigorously proved for certain
regimes. The exception is the wild multistability area where the
dynamics is very sensitive to variations of $N$.

The considered networks demonstrates new examples of remote
synchronization. These are quasiperiodic and chaotic regimes. For
these regimes the non-identity of the hub and the subordinate nodes is
not required to prevent their synchronization.

The most interesting situation occurs in the wild multistability
area. Similar type of dynamics was already reported for other systems,
in particular for coupled map lattices. We suggest to refer to it as
wild multistability because of its complexity and richness.  In fact
one can hardly predict here even a qualitative nature of the expected
solution. Each deviation in initial conditions and coupling strength
value is found to result in a new character of dynamics. This can be
either periodic, quasiperiodic, or chaotic regimes. Obviously a lot of
interesting problems arise in this connection. Since star-like motifs
are the basic building blocks of scale-free networks, the detailed
study of wild multistability area is very important for further
understanding of the dynamics of complex scale-free networks.

\section*{Acknowledgments}

This work was partially supported by a grant of the President of the
Russian Federation for leading scientific schools NSH-1726.2014.2
``Fundamental problems of nonlinear dynamics and their applications''

\bibliographystyle{apsrev4-1}
\bibliography{wildstars}

\begin{thebibliography}{31}%
\makeatletter
\providecommand \@ifxundefined [1]{%
 \@ifx{#1\undefined}
}%
\providecommand \@ifnum [1]{%
 \ifnum #1\expandafter \@firstoftwo
 \else \expandafter \@secondoftwo
 \fi
}%
\providecommand \@ifx [1]{%
 \ifx #1\expandafter \@firstoftwo
 \else \expandafter \@secondoftwo
 \fi
}%
\providecommand \natexlab [1]{#1}%
\providecommand \enquote  [1]{``#1''}%
\providecommand \bibnamefont  [1]{#1}%
\providecommand \bibfnamefont [1]{#1}%
\providecommand \citenamefont [1]{#1}%
\providecommand \href@noop [0]{\@secondoftwo}%
\providecommand \href [0]{\begingroup \@sanitize@url \@href}%
\providecommand \@href[1]{\@@startlink{#1}\@@href}%
\providecommand \@@href[1]{\endgroup#1\@@endlink}%
\providecommand \@sanitize@url [0]{\catcode `\\12\catcode `\$12\catcode
  `\&12\catcode `\#12\catcode `\^12\catcode `\_12\catcode `\%12\relax}%
\providecommand \@@startlink[1]{}%
\providecommand \@@endlink[0]{}%
\providecommand \url  [0]{\begingroup\@sanitize@url \@url }%
\providecommand \@url [1]{\endgroup\@href {#1}{\urlprefix }}%
\providecommand \urlprefix  [0]{URL }%
\providecommand \Eprint [0]{\href }%
\providecommand \doibase [0]{http://dx.doi.org/}%
\providecommand \selectlanguage [0]{\@gobble}%
\providecommand \bibinfo  [0]{\@secondoftwo}%
\providecommand \bibfield  [0]{\@secondoftwo}%
\providecommand \translation [1]{[#1]}%
\providecommand \BibitemOpen [0]{}%
\providecommand \bibitemStop [0]{}%
\providecommand \bibitemNoStop [0]{.\EOS\space}%
\providecommand \EOS [0]{\spacefactor3000\relax}%
\providecommand \BibitemShut  [1]{\csname bibitem#1\endcsname}%
\let\auto@bib@innerbib\@empty
\bibitem [{\citenamefont {Boccaletti}\ \emph {et~al.}(2006)\citenamefont
  {Boccaletti}, \citenamefont {Latora}, \citenamefont {Moreno}, \citenamefont
  {Chavez},\ and\ \citenamefont {Hwang}}]{Boccaletti2006175}%
  \BibitemOpen
  \bibfield  {author} {\bibinfo {author} {\bibfnamefont {S.}~\bibnamefont
  {Boccaletti}}, \bibinfo {author} {\bibfnamefont {V.}~\bibnamefont {Latora}},
  \bibinfo {author} {\bibfnamefont {Y.}~\bibnamefont {Moreno}}, \bibinfo
  {author} {\bibfnamefont {M.}~\bibnamefont {Chavez}}, \ and\ \bibinfo {author}
  {\bibfnamefont {D.-U.}\ \bibnamefont {Hwang}},\ }\href {\doibase
  10.1016/j.physrep.2005.10.009} {\bibfield  {journal} {\bibinfo  {journal}
  {Physics Reports}\ }\textbf {\bibinfo {volume} {424}},\ \bibinfo {pages} {175
  } (\bibinfo {year} {2006})}\BibitemShut {NoStop}%
\bibitem [{\citenamefont {Wang}(2002)}]{CxNetwTopDynSyn2002}%
  \BibitemOpen
  \bibfield  {author} {\bibinfo {author} {\bibfnamefont {X.~F.}\ \bibnamefont
  {Wang}},\ }\href {\doibase 10.1142/S0218127402004802} {\bibfield  {journal}
  {\bibinfo  {journal} {International Journal of Bifurcation and Chaos}\
  }\textbf {\bibinfo {volume} {12}},\ \bibinfo {pages} {885} (\bibinfo {year}
  {2002})}\BibitemShut {NoStop}%
\bibitem [{\citenamefont {Barabási}\ \emph {et~al.}(2000)\citenamefont
  {Barabási}, \citenamefont {Albert},\ and\ \citenamefont
  {Jeong}}]{ScaleFreeNetw}%
  \BibitemOpen
  \bibfield  {author} {\bibinfo {author} {\bibfnamefont {A.-L.}\ \bibnamefont
  {Barabási}}, \bibinfo {author} {\bibfnamefont {R.}~\bibnamefont {Albert}}, \
  and\ \bibinfo {author} {\bibfnamefont {H.}~\bibnamefont {Jeong}},\ }\href
  {\doibase 10.1016/S0378-4371(00)00018-2} {\bibfield  {journal} {\bibinfo
  {journal} {Physica A}\ }\textbf {\bibinfo {volume} {281}},\ \bibinfo {pages}
  {69 } (\bibinfo {year} {2000})}\BibitemShut {NoStop}%
\bibitem [{\citenamefont {Arenas}\ \emph {et~al.}(2008)\citenamefont {Arenas},
  \citenamefont {Díaz-Guilera}, \citenamefont {Kurths}, \citenamefont
  {Moreno},\ and\ \citenamefont {Zhou}}]{Arenas200893}%
  \BibitemOpen
  \bibfield  {author} {\bibinfo {author} {\bibfnamefont {A.}~\bibnamefont
  {Arenas}}, \bibinfo {author} {\bibfnamefont {A.}~\bibnamefont
  {Díaz-Guilera}}, \bibinfo {author} {\bibfnamefont {J.}~\bibnamefont
  {Kurths}}, \bibinfo {author} {\bibfnamefont {Y.}~\bibnamefont {Moreno}}, \
  and\ \bibinfo {author} {\bibfnamefont {C.}~\bibnamefont {Zhou}},\ }\href
  {\doibase 10.1016/j.physrep.2008.09.002} {\bibfield  {journal} {\bibinfo
  {journal} {Physics Reports}\ }\textbf {\bibinfo {volume} {469}},\ \bibinfo
  {pages} {93 } (\bibinfo {year} {2008})}\BibitemShut {NoStop}%
\bibitem [{\citenamefont {Osipov}\ \emph {et~al.}(2007)\citenamefont {Osipov},
  \citenamefont {Kurths},\ and\ \citenamefont
  {Zhou}}]{osipov2007synchronization}%
  \BibitemOpen
  \bibfield  {author} {\bibinfo {author} {\bibfnamefont {G.~V.}\ \bibnamefont
  {Osipov}}, \bibinfo {author} {\bibfnamefont {J.}~\bibnamefont {Kurths}}, \
  and\ \bibinfo {author} {\bibfnamefont {C.}~\bibnamefont {Zhou}},\ }\href@noop
  {} {\emph {\bibinfo {title} {Synchronization in oscillatory networks}}}\
  (\bibinfo  {publisher} {Springer Science \& Business Media},\ \bibinfo {year}
  {2007})\BibitemShut {NoStop}%
\bibitem [{\citenamefont {Golubitsky}\ and\ \citenamefont
  {Stewart}(2015)}]{golubitsky2015recent}%
  \BibitemOpen
  \bibfield  {author} {\bibinfo {author} {\bibfnamefont {M.}~\bibnamefont
  {Golubitsky}}\ and\ \bibinfo {author} {\bibfnamefont {I.}~\bibnamefont
  {Stewart}},\ }\href {\doibase 10.1063/1.4918595} {\bibfield  {journal}
  {\bibinfo  {journal} {Chaos: An Interdisciplinary Journal of Nonlinear
  Science}\ }\textbf {\bibinfo {volume} {25}},\ \bibinfo {pages} {097612}
  (\bibinfo {year} {2015})}\BibitemShut {NoStop}%
\bibitem [{\citenamefont {Belykh}\ \emph {et~al.}(2005)\citenamefont {Belykh},
  \citenamefont {Hasler}, \citenamefont {Lauret},\ and\ \citenamefont
  {Nijmeijer}}]{SyncGraphTopol2005}%
  \BibitemOpen
  \bibfield  {author} {\bibinfo {author} {\bibfnamefont {I.}~\bibnamefont
  {Belykh}}, \bibinfo {author} {\bibfnamefont {M.}~\bibnamefont {Hasler}},
  \bibinfo {author} {\bibfnamefont {M.}~\bibnamefont {Lauret}}, \ and\ \bibinfo
  {author} {\bibfnamefont {H.}~\bibnamefont {Nijmeijer}},\ }\href {\doibase
  10.1142/S0218127405014143} {\bibfield  {journal} {\bibinfo  {journal}
  {International Journal of Bifurcation and Chaos}\ }\textbf {\bibinfo {volume}
  {15}},\ \bibinfo {pages} {3423} (\bibinfo {year} {2005})}\BibitemShut
  {NoStop}%
\bibitem [{\citenamefont {Arenas}\ \emph {et~al.}(2006)\citenamefont {Arenas},
  \citenamefont {D\'{i}az-Guilera},\ and\ \citenamefont
  {P\'erez-Vicente}}]{arenas2006synchronization}%
  \BibitemOpen
  \bibfield  {author} {\bibinfo {author} {\bibfnamefont {A.}~\bibnamefont
  {Arenas}}, \bibinfo {author} {\bibfnamefont {A.}~\bibnamefont
  {D\'{i}az-Guilera}}, \ and\ \bibinfo {author} {\bibfnamefont {C.~J.}\
  \bibnamefont {P\'erez-Vicente}},\ }\href {\doibase
  10.1103/PhysRevLett.96.114102} {\bibfield  {journal} {\bibinfo  {journal}
  {Phys. Rev. Lett.}\ }\textbf {\bibinfo {volume} {96}},\ \bibinfo {pages}
  {114102} (\bibinfo {year} {2006})}\BibitemShut {NoStop}%
\bibitem [{\citenamefont {Wang}\ and\ \citenamefont
  {Chen}(2002{\natexlab{a}})}]{wang2002synchronization}%
  \BibitemOpen
  \bibfield  {author} {\bibinfo {author} {\bibfnamefont {X.~F.}\ \bibnamefont
  {Wang}}\ and\ \bibinfo {author} {\bibfnamefont {G.}~\bibnamefont {Chen}},\
  }\href {\doibase 10.1109/81.974874} {\bibfield  {journal} {\bibinfo
  {journal} {Circuits and Systems I: Fundamental Theory and Applications, IEEE
  Transactions on}\ }\textbf {\bibinfo {volume} {49}},\ \bibinfo {pages} {54}
  (\bibinfo {year} {2002}{\natexlab{a}})}\BibitemShut {NoStop}%
\bibitem [{\citenamefont {Fan}\ and\ \citenamefont
  {Wang}(2005)}]{SyncScaleFree2005}%
  \BibitemOpen
  \bibfield  {author} {\bibinfo {author} {\bibfnamefont {J.}~\bibnamefont
  {Fan}}\ and\ \bibinfo {author} {\bibfnamefont {X.~F.}\ \bibnamefont {Wang}},\
  }\href {\doibase 10.1016/j.physa.2004.09.016} {\bibfield  {journal} {\bibinfo
   {journal} {Physica A: Statistical Mechanics and its Applications}\ }\textbf
  {\bibinfo {volume} {349}},\ \bibinfo {pages} {443 } (\bibinfo {year}
  {2005})}\BibitemShut {NoStop}%
\bibitem [{\citenamefont {Jalan}\ and\ \citenamefont
  {Amritkar}(2003)}]{JalanAmritkar2003}%
  \BibitemOpen
  \bibfield  {author} {\bibinfo {author} {\bibfnamefont {S.}~\bibnamefont
  {Jalan}}\ and\ \bibinfo {author} {\bibfnamefont {R.~E.}\ \bibnamefont
  {Amritkar}},\ }\href {\doibase 10.1103/PhysRevLett.90.014101} {\bibfield
  {journal} {\bibinfo  {journal} {Phys. Rev. Lett.}\ }\textbf {\bibinfo
  {volume} {90}},\ \bibinfo {pages} {014101} (\bibinfo {year}
  {2003})}\BibitemShut {NoStop}%
\bibitem [{\citenamefont {Jalan}\ \emph {et~al.}(2005)\citenamefont {Jalan},
  \citenamefont {Amritkar},\ and\ \citenamefont {Hu}}]{JalanAmritkar2005}%
  \BibitemOpen
  \bibfield  {author} {\bibinfo {author} {\bibfnamefont {S.}~\bibnamefont
  {Jalan}}, \bibinfo {author} {\bibfnamefont {R.~E.}\ \bibnamefont {Amritkar}},
  \ and\ \bibinfo {author} {\bibfnamefont {C.-K.}\ \bibnamefont {Hu}},\ }\href
  {\doibase 10.1103/PhysRevE.72.016211} {\bibfield  {journal} {\bibinfo
  {journal} {Phys. Rev. E}\ }\textbf {\bibinfo {volume} {72}},\ \bibinfo
  {pages} {016211} (\bibinfo {year} {2005})}\BibitemShut {NoStop}%
\bibitem [{\citenamefont {Jalan}\ \emph {et~al.}(2015)\citenamefont {Jalan},
  \citenamefont {Singh}, \citenamefont {Acharyya},\ and\ \citenamefont
  {Kurths}}]{ImpLeadClSync2015}%
  \BibitemOpen
  \bibfield  {author} {\bibinfo {author} {\bibfnamefont {S.}~\bibnamefont
  {Jalan}}, \bibinfo {author} {\bibfnamefont {A.}~\bibnamefont {Singh}},
  \bibinfo {author} {\bibfnamefont {S.}~\bibnamefont {Acharyya}}, \ and\
  \bibinfo {author} {\bibfnamefont {J.}~\bibnamefont {Kurths}},\ }\href
  {\doibase 10.1103/PhysRevE.91.022901} {\bibfield  {journal} {\bibinfo
  {journal} {Phys. Rev. E}\ }\textbf {\bibinfo {volume} {91}},\ \bibinfo
  {pages} {022901} (\bibinfo {year} {2015})}\BibitemShut {NoStop}%
\bibitem [{\citenamefont {Gambuzza}\ \emph {et~al.}(2013)\citenamefont
  {Gambuzza}, \citenamefont {Cardillo}, \citenamefont {Fiasconaro},
  \citenamefont {Gomez-Gardenes},\ and\ \citenamefont {Frasca}}]{RemSyn2}%
  \BibitemOpen
  \bibfield  {author} {\bibinfo {author} {\bibfnamefont {L.~V.}\ \bibnamefont
  {Gambuzza}}, \bibinfo {author} {\bibfnamefont {A.}~\bibnamefont {Cardillo}},
  \bibinfo {author} {\bibfnamefont {A.}~\bibnamefont {Fiasconaro}}, \bibinfo
  {author} {\bibfnamefont {L.~F.~J.}\ \bibnamefont {Gomez-Gardenes}}, \ and\
  \bibinfo {author} {\bibfnamefont {M.}~\bibnamefont {Frasca}},\ }\href
  {\doibase 10.1063/1.4824312} {\bibfield  {journal} {\bibinfo  {journal}
  {Chaos}\ }\textbf {\bibinfo {volume} {23}},\ \bibinfo {pages} {043103}
  (\bibinfo {year} {2013})}\BibitemShut {NoStop}%
\bibitem [{\citenamefont {Wang}\ and\ \citenamefont
  {Zhang}(2010)}]{Wang20101464}%
  \BibitemOpen
  \bibfield  {author} {\bibinfo {author} {\bibfnamefont {J.}~\bibnamefont
  {Wang}}\ and\ \bibinfo {author} {\bibfnamefont {Y.}~\bibnamefont {Zhang}},\
  }\href {\doibase 10.1016/j.physleta.2010.01.042} {\bibfield  {journal}
  {\bibinfo  {journal} {Physics Letters A}\ }\textbf {\bibinfo {volume}
  {374}},\ \bibinfo {pages} {1464 } (\bibinfo {year} {2010})}\BibitemShut
  {NoStop}%
\bibitem [{\citenamefont {Wang}\ and\ \citenamefont
  {Chen}(2002{\natexlab{b}})}]{wang2002pinning}%
  \BibitemOpen
  \bibfield  {author} {\bibinfo {author} {\bibfnamefont {X.~F.}\ \bibnamefont
  {Wang}}\ and\ \bibinfo {author} {\bibfnamefont {G.}~\bibnamefont {Chen}},\
  }\href {\doibase http://dx.doi.org/10.1016/S0378-4371(02)00772-0} {\bibfield
  {journal} {\bibinfo  {journal} {Physica A: Statistical Mechanics and its
  Applications}\ }\textbf {\bibinfo {volume} {310}},\ \bibinfo {pages} {521 }
  (\bibinfo {year} {2002}{\natexlab{b}})}\BibitemShut {NoStop}%
\bibitem [{\citenamefont {Kuptsov}\ and\ \citenamefont
  {Parlitz}(2012)}]{CLV2012}%
  \BibitemOpen
  \bibfield  {author} {\bibinfo {author} {\bibfnamefont {P.~V.}\ \bibnamefont
  {Kuptsov}}\ and\ \bibinfo {author} {\bibfnamefont {U.}~\bibnamefont
  {Parlitz}},\ }\href {\doibase 10.1007/s00332-012-9126-5} {\bibfield
  {journal} {\bibinfo  {journal} {J. Nonlinear Sci.}\ }\textbf {\bibinfo
  {volume} {22}},\ \bibinfo {pages} {727} (\bibinfo {year} {2012})}\BibitemShut
  {NoStop}%
\bibitem [{\citenamefont {Kuptsov}\ and\ \citenamefont
  {Kuptsova}(2014)}]{NWL2014}%
  \BibitemOpen
  \bibfield  {author} {\bibinfo {author} {\bibfnamefont {P.~V.}\ \bibnamefont
  {Kuptsov}}\ and\ \bibinfo {author} {\bibfnamefont {A.~V.}\ \bibnamefont
  {Kuptsova}},\ }\href {\doibase 10.1103/PhysRevE.90.032901} {\bibfield
  {journal} {\bibinfo  {journal} {Phys. Rev. E}\ }\textbf {\bibinfo {volume}
  {90}},\ \bibinfo {pages} {032901} (\bibinfo {year} {2014})}\BibitemShut
  {NoStop}%
\bibitem [{\citenamefont {Roxin}\ \emph {et~al.}(2005)\citenamefont {Roxin},
  \citenamefont {Brunel},\ and\ \citenamefont {Hansel}}]{Roxin2005}%
  \BibitemOpen
  \bibfield  {author} {\bibinfo {author} {\bibfnamefont {A.}~\bibnamefont
  {Roxin}}, \bibinfo {author} {\bibfnamefont {N.}~\bibnamefont {Brunel}}, \
  and\ \bibinfo {author} {\bibfnamefont {D.}~\bibnamefont {Hansel}},\ }\href
  {\doibase 10.1103/PhysRevLett.94.238103} {\bibfield  {journal} {\bibinfo
  {journal} {Phys. Rev. Lett.}\ }\textbf {\bibinfo {volume} {94}},\ \bibinfo
  {pages} {238103} (\bibinfo {year} {2005})}\BibitemShut {NoStop}%
\bibitem [{\citenamefont {Angeli}\ \emph {et~al.}(2004)\citenamefont {Angeli},
  \citenamefont {Ferrell},\ and\ \citenamefont {Sontag}}]{Angeli2004}%
  \BibitemOpen
  \bibfield  {author} {\bibinfo {author} {\bibfnamefont {D.}~\bibnamefont
  {Angeli}}, \bibinfo {author} {\bibfnamefont {J.~E.}\ \bibnamefont {Ferrell}},
  \ and\ \bibinfo {author} {\bibfnamefont {E.~D.}\ \bibnamefont {Sontag}},\
  }\href {\doibase 10.1073/pnas.0308265100} {\bibfield  {journal} {\bibinfo
  {journal} {Proceedings of the National Academy of Sciences of the United
  States of America}\ }\textbf {\bibinfo {volume} {101}},\ \bibinfo {pages}
  {1822} (\bibinfo {year} {2004})}\BibitemShut {NoStop}%
\bibitem [{\citenamefont {Angeli}(2009)}]{Angeli2009398}%
  \BibitemOpen
  \bibfield  {author} {\bibinfo {author} {\bibfnamefont {D.}~\bibnamefont
  {Angeli}},\ }\href {\doibase 10.3166/ejc.15.398-406} {\bibfield  {journal}
  {\bibinfo  {journal} {European Journal of Control}\ }\textbf {\bibinfo
  {volume} {15}},\ \bibinfo {pages} {398 } (\bibinfo {year}
  {2009})}\BibitemShut {NoStop}%
\bibitem [{\citenamefont {Feudel}(2008)}]{Feudel2008}%
  \BibitemOpen
  \bibfield  {author} {\bibinfo {author} {\bibfnamefont {U.}~\bibnamefont
  {Feudel}},\ }\href {\doibase 10.1142/S0218127408021233} {\bibfield  {journal}
  {\bibinfo  {journal} {International Journal of Bifurcation and Chaos}\
  }\textbf {\bibinfo {volume} {18}},\ \bibinfo {pages} {1607} (\bibinfo {year}
  {2008})}\BibitemShut {NoStop}%
\bibitem [{\citenamefont {Pisarchik}\ and\ \citenamefont
  {Feudel}(2014)}]{Pisarchik2014167}%
  \BibitemOpen
  \bibfield  {author} {\bibinfo {author} {\bibfnamefont {A.~N.}\ \bibnamefont
  {Pisarchik}}\ and\ \bibinfo {author} {\bibfnamefont {U.}~\bibnamefont
  {Feudel}},\ }\href {\doibase 10.1016/j.physrep.2014.02.007} {\bibfield
  {journal} {\bibinfo  {journal} {Physics Reports}\ }\textbf {\bibinfo {volume}
  {540}},\ \bibinfo {pages} {167 } (\bibinfo {year} {2014})},\ \bibinfo {note}
  {control of multistability}\BibitemShut {NoStop}%
\bibitem [{\citenamefont {Pecora}(1998)}]{Pecora98}%
  \BibitemOpen
  \bibfield  {author} {\bibinfo {author} {\bibfnamefont {L.~M.}\ \bibnamefont
  {Pecora}},\ }\href {\doibase 10.1103/PhysRevE.58.347} {\bibfield  {journal}
  {\bibinfo  {journal} {Phys. Rev. E}\ }\textbf {\bibinfo {volume} {58}},\
  \bibinfo {pages} {347} (\bibinfo {year} {1998})}\BibitemShut {NoStop}%
\bibitem [{\citenamefont {Ma}\ \emph {et~al.}(2008)\citenamefont {Ma},
  \citenamefont {Zhang}, \citenamefont {Wang},\ and\ \citenamefont
  {Liu}}]{ClustSyncStar2008}%
  \BibitemOpen
  \bibfield  {author} {\bibinfo {author} {\bibfnamefont {Z.}~\bibnamefont
  {Ma}}, \bibinfo {author} {\bibfnamefont {G.}~\bibnamefont {Zhang}}, \bibinfo
  {author} {\bibfnamefont {Y.}~\bibnamefont {Wang}}, \ and\ \bibinfo {author}
  {\bibfnamefont {Z.}~\bibnamefont {Liu}},\ }\href {\doibase
  10.1088/1751-8113/41/15/155101} {\bibfield  {journal} {\bibinfo  {journal}
  {Journal of Physics A: Mathematical and Theoretical}\ }\textbf {\bibinfo
  {volume} {41}},\ \bibinfo {pages} {155101} (\bibinfo {year}
  {2008})}\BibitemShut {NoStop}%
\bibitem [{\citenamefont {Bergner}\ \emph {et~al.}(2012)\citenamefont
  {Bergner}, \citenamefont {Frasca}, \citenamefont {Sciuto}, \citenamefont
  {Buscarino}, \citenamefont {Ngamga}, \citenamefont {Fortuna},\ and\
  \citenamefont {Kurths}}]{RemSyn1}%
  \BibitemOpen
  \bibfield  {author} {\bibinfo {author} {\bibfnamefont {A.}~\bibnamefont
  {Bergner}}, \bibinfo {author} {\bibfnamefont {M.}~\bibnamefont {Frasca}},
  \bibinfo {author} {\bibfnamefont {G.}~\bibnamefont {Sciuto}}, \bibinfo
  {author} {\bibfnamefont {A.}~\bibnamefont {Buscarino}}, \bibinfo {author}
  {\bibfnamefont {E.~J.}\ \bibnamefont {Ngamga}}, \bibinfo {author}
  {\bibfnamefont {L.}~\bibnamefont {Fortuna}}, \ and\ \bibinfo {author}
  {\bibfnamefont {J.}~\bibnamefont {Kurths}},\ }\href {\doibase
  10.1103/PhysRevE.85.026208} {\bibfield  {journal} {\bibinfo  {journal} {Phys.
  Rev. E}\ }\textbf {\bibinfo {volume} {85}},\ \bibinfo {pages} {026208}
  (\bibinfo {year} {2012})}\BibitemShut {NoStop}%
\bibitem [{\citenamefont {Hénon}(1976)}]{Henon1976}%
  \BibitemOpen
  \bibfield  {author} {\bibinfo {author} {\bibfnamefont {M.}~\bibnamefont
  {Hénon}},\ }\href {\doibase 10.1007/BF01608556} {\bibfield  {journal}
  {\bibinfo  {journal} {Communications in Mathematical Physics}\ }\textbf
  {\bibinfo {volume} {50}},\ \bibinfo {pages} {69} (\bibinfo {year}
  {1976})}\BibitemShut {NoStop}%
\bibitem [{\citenamefont {Politi}\ and\ \citenamefont
  {Torcini}(1992)}]{PolTor92a}%
  \BibitemOpen
  \bibfield  {author} {\bibinfo {author} {\bibfnamefont {A.}~\bibnamefont
  {Politi}}\ and\ \bibinfo {author} {\bibfnamefont {A.}~\bibnamefont
  {Torcini}},\ }\href {\doibase 10.1063/1.165871} {\bibfield  {journal}
  {\bibinfo  {journal} {Chaos}\ }\textbf {\bibinfo {volume} {2}},\ \bibinfo
  {pages} {293} (\bibinfo {year} {1992})}\BibitemShut {NoStop}%
\bibitem [{\citenamefont {Feudel}\ \emph {et~al.}(2006)\citenamefont {Feudel},
  \citenamefont {Kuznetsov},\ and\ \citenamefont {Pikovsky}}]{SNA}%
  \BibitemOpen
  \bibfield  {author} {\bibinfo {author} {\bibfnamefont {U.}~\bibnamefont
  {Feudel}}, \bibinfo {author} {\bibfnamefont {S.}~\bibnamefont {Kuznetsov}}, \
  and\ \bibinfo {author} {\bibfnamefont {A.}~\bibnamefont {Pikovsky}},\
  }\href@noop {} {\emph {\bibinfo {title} {Strange nonchaotic attractors}}}\
  (\bibinfo  {publisher} {World Scientific},\ \bibinfo {year}
  {2006})\BibitemShut {NoStop}%
\bibitem [{\citenamefont {Pecora}\ and\ \citenamefont {Carroll}(1998)}]{MSF98}%
  \BibitemOpen
  \bibfield  {author} {\bibinfo {author} {\bibfnamefont {L.~M.}\ \bibnamefont
  {Pecora}}\ and\ \bibinfo {author} {\bibfnamefont {T.~L.}\ \bibnamefont
  {Carroll}},\ }\href {\doibase 10.1103/PhysRevLett.80.2109} {\bibfield
  {journal} {\bibinfo  {journal} {Phys. Rev. Lett.}\ }\textbf {\bibinfo
  {volume} {80}},\ \bibinfo {pages} {2109} (\bibinfo {year}
  {1998})}\BibitemShut {NoStop}%
\bibitem [{\citenamefont {Milnor}(2004)}]{milnor2004concept}%
  \BibitemOpen
  \bibfield  {author} {\bibinfo {author} {\bibfnamefont {J.}~\bibnamefont
  {Milnor}},\ }in\ \href@noop {} {\emph {\bibinfo {booktitle} {The Theory of
  Chaotic Attractors}}}\ (\bibinfo  {publisher} {Springer},\ \bibinfo {year}
  {2004})\ pp.\ \bibinfo {pages} {243--264}\BibitemShut {NoStop}%
\end{thebibliography}%

\end{document}